\documentclass[slac_one]{revtex4}
\usepackage{graphicx}
\usepackage{fancyhdr}
\begin{document}

\title{Top quark theory review for the Tevatron, LHC, and ILC}
\author{D. Wackeroth}
\affiliation{University at Buffalo, The State University of New York, Buffalo, NY 14260, USA}
\begin{abstract}
  I will briefly review the status of higher-order calculations for top-quark
  observables, comment on the need for improvements, discuss some of the recent
  theoretical advances, and present a few examples to highlight the role of
  top-quark observables in searches for signals of physics beyond the Standard
  Model.
\end{abstract}
\maketitle

\thispagestyle{fancy}

\section{THE MANY FACETS OF TOP QUARK PHYSICS} 

The study of top-quark properties and dynamics provides a unique
window to the mechanism of electroweak symmetry breaking (EWSB).  The
large mass of the top quark suggests that it plays a special role in
EWSB, and that new physics connected to EWSB may be found first
through precision studies of top-quark observables. Deviations of
experimental measurements from the SM predictions, including
electroweak (EW) and QCD corrections, could indicate new non-standard
top production or decay mechanisms. Since the top quark immediately
decays before it hadronizes or flips its spin, it provides an
excellent testing ground for perturbative QCD. Moreover, information
about spin correlation and polarization, imprinted by the production
process, is preserved, and can be measured in angular distributions of
top-decay products, providing another way to search for deviations
from the SM expectation.  The precise measurement of the top-quark
mass ($m_t$) allows for improved bounds on the mass of the Standard
Model (SM) Higgs boson ($M_H$), which is presently constrained to be
smaller than 185~GeV ($95\%$ C.L.)~\cite{Alcaraz:2007ri}.  Measuring
precisely the properties of the top quark and studying its dynamics
therefore is an important goal at the Tevatron Run~II, LHC and ILC. To
fully exploit the potential of these colliders for precision top-quark
physics, it is crucial that predictions for top-quark observables are
under theoretical control and include higher-order corrections within
the SM and beyond. In the following, I will briefly describe some
theory aspects of top-quark physics relevant to the Tevatron, LHC and
ILC.  For detailed reviews of theoretical and experimental results
please see, e.g.,
Refs.~\cite{Bernreuther:2008ju,Kehoe:2007px,Gerber:2007xk,Abe:2001nqa,Beneke:2000hk}
(as well as presentations in this session).

\section{TOTAL PRODUCTION CROSS SECTIONS: $t\bar t$ AND SINGLE TOP}

The total top-pair production cross section ($\sigma_{t\bar t}$) is
presently measured at the Tevatron~\footnote{See www-cdf.fnal.gov and
www-d0.fnal.gov for most recent results from the CDF and D0
collaborations, respectively.}  with a relative uncertainty of
$\Delta\sigma_{t\bar t}/\sigma_{t\bar t}=9(11)\%$
(CDF~\cite{ttcdf}(D0~\cite{Abazov:2008gc})) (with ${\cal L}=2.8(0.9)
\; {\rm fb}^{-1}$).  While QCD
predictions~\cite{Moch:2008qy,Cacciari:2008zb,Kidonakis:2008mu} and
measurements of $\sigma_{t\bar t}$ agree within their respective
uncertainties, recent detailed
studies~\cite{Moch:2008qy,Cacciari:2008zb,Kidonakis:2008mu} of the
theoretical uncertainties of presently available state-of-the-art QCD
calculations show that further theoretical improvements will be
necessary in order to match (or better exceed) the anticipated future
experimental precision, as illustrated in Table~\ref{ttbartheory}. For
instance at the LHC, the goal is to measure $\sigma_{t\bar t}$
ultimately with a relative uncertainty of $\approx 5\%$. All presently
available QCD predictions for the total cross sections of the strong
$t \bar t$ production processes, $q\bar q \to t\bar t$ and $gg \to
t\bar t$, include the complete fixed-order next-to-leading-order (NLO)
corrections~\cite{ttqcdfixed} and next-to-leading-logarithmic (NLL)
contributions due to soft gluon radiation at the $t\bar t$ threshold
resummed to all orders~\cite{ttqcdresummed}. However, as illustrated
in Table~\ref{ttbartheory} and also pointed out in
Ref.~\cite{Bern:2008ef}, the next-to-next-to-leading-order (NNLO) QCD
corrections are needed.  Without having a full NNLO calculation at
hand, recent advances aim to extract partial NNLO
contributions~\cite{Moch:2008qy,Kidonakis:2008mu} (labeled NNLO
approx. in Table~\ref{ttbartheory}) from an expansion of the
threshold-resummed results at next-to-next-to-leading logarithmic
(NNLL) accuracy. The residual theoretical uncertainty of these
predictions due to missing higher-order corrections is estimated by
varying the renormalization and factorization scales as shown in
Table~\ref{ttbartheory}.  The PDF uncertainty is also provided for two
sets of PDFs, MRST2006nnlo~\cite{Martin:2007bv} and CTEQ6.5 (or
CTEQ6.6M)~\cite{Nadolsky:2008zw}.  For a complete fixed-order NNLO
calculation of $\sigma_{t\bar t}$ one needs the leading-order $2 \to
4$ parton scattering process, the $2 \to 3$ process at NLO, the NLO $2
\to 2$ process squared, virtual 2-loop corrections to the $2 \to 2$
process, a treatment of soft and collinear singularities at NNLO (see,
e.g., Refs.~\cite{GehrmannDeRidder:2005cm,Daleo:2006xa}), and last but
not least NNLO PDFs.  The 2-loop virtual QCD corrections to $q\bar q
\to t\bar t$ have been evaluated numerically in
Ref.~\cite{Czakon:2008zk} .  An analytic result for 2-loop fermion
loops in $q\bar q \to t\bar t$ has been provided as
well~\cite{Bonciani:2008az}.  Results for two-loop corrections to both
$q\bar q \to t \bar t$ and $gg\to t\bar t$ have been obtained in the
limit $s,|t|,|u| \gg m_Q^2$ in
Refs.~\cite{Czakon:2007ej,Czakon:2007wk}.  NNLO ${\cal O}(\alpha_s^4)$
one-loop squared corrections to both production processes $q \bar q
\to t\bar t$~\cite{Korner:2008bn} and $gg \to t\bar
t$~\cite{Kniehl:2008fd,Anastasiou:2008vd} with full mass dependence
have been finalized recently. The $t \bar t+$jet cross section at NLO
QCD is provided in Refs.~\cite{Dittmaier:2007wz,Dittmaier:2008uj}.
For a complete consistent NNLO calculation NNLO PDFs are
needed. MRST2006~\cite{Martin:2007bv} uses NNLO evolution
kernels~\cite{Vogt:2004mw,Moch:2004pa} and NNLO QCD predictions for
Drell-Yan cross sections~\cite{Anastasiou:2003ds}.
SM EW radiative corrections to $\sigma_{t\bar t}$ have been studied at
NLO in
Refs.~\cite{Beenakker:1993yr,Kao:1997bs,Kuhn:2005it,Bernreuther:2005is,Moretti:2006nf,Moretti:2006ea,Kuhn:2006vh,Hollik:2007sw,Bernreuther:2008aw}.
While they are known to only have a small impact on $\sigma_{t\bar t}$
($\approx 1-2 \%$), EW corrections can significantly affect top-quark
distributions at high energies due to the occurrence of large EW
Sudakov-like logarithms~\cite{Kuhn:2006vh}. SUSY
EW~\cite{Yang:1996dma,Yang:1995hq,Kim:1996nza,Hollik:1997hm,Ross:2007ez} and SUSY
QCD~\cite{Alam:1996mh,Sullivan:1996ry,Zhou:1997fw,Yu:1998xv,Berge:2007dz,Ross:2007ez}
one-loop corrections have been calculated for both $q\bar q \to t \bar
t$ and $gg \to t\bar t$. Supersymmetric particles in loops can affect
$\sigma_{t\bar t}$ by up to $\approx 6\%$~\cite{Ross:2007ez}.  More
interesting observables to observe such effects are kinematic
distributions and asymmetries as will be discussed in
Section~\ref{sec:dynamics}.

\begin{table}[t]
\begin{center}
\caption{Theoretical uncertainties of state-of-the-art QCD predictions for $\sigma_{t\bar t}$
at the Tevatron ($\sqrt{s}=$1.96 TeV) and the LHC (with $m_t=171$ GeV) due to
scale dependence and PDF uncertainties. The total uncertainties have been
calculated by adding the scale and PDF uncertainties (and in case of
Ref.~\cite{Kidonakis:2008mu} also the kinematic uncertainties) in quadrature.} 
\begin{tabular}{|c|c|c|c||c|c|c|}\hline
\textbf{Tevatron}& \multicolumn{3}{|c|}{\textbf{MRST2006nnlo}} & \multicolumn{3}{|c|}{\textbf{CTEQ6.5/6.6M}} \\ \hline
$\Delta\sigma_{t\bar t}/\sigma_{t\bar t}[\%]$                     & scale & PDF & total & scale & PDF & total  \\ \hline
NLO+NLL~\cite{Cacciari:2008zb}& +4,-7& 3 & +5,-7 & +4,-7 & +7,-5& +8,-8\\
NNLO approx.~\cite{Moch:2008qy}& 3 & 3 & 6 & 3 & 6 & 8  \\
NNLO approx.~\cite{Kidonakis:2008mu} & +0.4,-3 & +3,-2& +5,-6 & +0.4,-3 & +7,-5& +8,-7  \\ \hline 
\textbf{LHC}& \multicolumn{3}{|c|}{\textbf{MRST2006nnlo}} & \multicolumn{3}{|c|}{\textbf{CTEQ6.5/6.6M}} \\ \hline
$\Delta\sigma_{t\bar t}/\sigma_{t\bar t}[\%]$                     & scale & PDF & total & scale & PDF & total \\ \hline
NLO+NLL~\cite{Cacciari:2008zb} & 9& 1& 9 & 9 & 3 & 10 \\
NNLO approx.~\cite{Moch:2008qy} & 3 & 2 & 4 & 3& 4 & 6  \\
NNLO approx.~\cite{Kidonakis:2008mu} & +8,-5& 1 & +8,-5 & +8,-5& 3& +8,-6  \\ \hline 
\end{tabular}
\label{ttbartheory}
\end{center}
\end{table}

The total cross section for single top-quark production ($\sigma_t$)
has been measured only recently at the
Tevatron~\cite{Aaltonen:2008sy,Abazov:2006gd}, providing a first
direct measurement of $V_{tb}$.  Single top production can play an
important role at the LHC in identifying and discriminating between
different new physics models~\cite{Tait:2000sh,Cao:2007ea}, since they
can have different, model-specific effects on the three production
processes, $s$-channel and $t$-channel (dominant at both the Tevatron
and the LHC) $tq$ production, and associated $tW$ production. The NLO QCD
corrections to these processes are known for both an on-shell top
quark~\cite{singletfixed} and including the top-quark
decay~\cite{singletfixedwdecay}. Improved QCD predictions for $\sigma_t$
also include NLL threshold resummation
effects~\cite{Mrenna:1997wp,Kidonakis:2006bu,Kidonakis:2007ej}, resulting in a residual theoretical
uncertainty of $\Delta \sigma_t/\sigma_t \approx 4-5\%$ as shown in
Table~\ref{singlettheory}.  Complete one-loop EW corrections
to the $t$-channel production process have been studied in Ref.~\cite{Beccaria:2008av}
and only have a modest effect on $\sigma_t$ (a few percent at the
LHC).  Genuine SUSY effects, at least within mSUGRA, affect $\sigma_t$
($t$-channel) at the LHC by at most $1\%$~\cite{Zhang:2006cx,Beccaria:2008av}. 

\begin{table}[t]
\begin{center}
\caption{Predictions for $\sigma_t$ at the Tevatron ($\sqrt{s}$=1.96 TeV) and 
the LHC (with $m_t=171.4$ GeV and MRST2004nnlo) including scale, 
PDF, and $\Delta m_t$
uncertainties~\cite{Kidonakis:2006bu,Kidonakis:2007ej}. The NNNLO approximate
results include the exact NLO QCD cross sections and an expansion of NLL resummed soft gluon corrections
through NNNLO. For comparison, the NNLO
approximate result for $\sigma_{t\bar t}$
from Ref.~\cite{Kidonakis:2008mu} including scale, kinematics and PDF uncertainties is also provided.} 
\begin{tabular}{|c|c|c|} \hline
$\sigma_t$ [pb] & \textbf{Tevatron} & \textbf{LHC} \\ \hline 
$t$-channel (NNNLO approx.)& $1.15\pm 0.07$ & $150 \pm 6$ \\ \hline 
$s$-channel (NNNLO approx.) & $0.54 \pm 0.04$ & $7.8+0.7-0.6$ \\ \hline 
$tW$ mode (NNLO/NNNLO approx.) &   $0.14 \pm 0.03$ & $43.5 \pm 4.8$ \\ \hline 
$\sigma_{t\bar t}$ (NNLO approx., $m_t$=172 GeV)[pb]& $7.80+0.39-0.45$ & $968+80-52$ \\ \hline \hline
\end{tabular}
\label{singlettheory}
\end{center}
\end{table}

\section{TOP QUARK MASS}

The impressively precise top-mass measurement at the Tevatron ($\Delta
m_t^{exp}/m_t^{exp}= 0.7\%$~\cite{mtop:2008vn}) and future
high-precision measurements at the LHC and the ILC require a
theoretically stable mass definition in a suitable renormalization
scheme, so that $m_t^{exp}$ can be related with equally high precision
to the $\overline{{\rm MS}}$ or on-shell masses used in global fits to
EW precision observables. Higgs-mass bounds that are extracted from
such fits strongly depend on the value of $m_t$. For instance, a
change in $m_t$ of 2~GeV shifts the central value of $M_H$ by about
15\%~\cite{mtmhshift}.  At the ILC, a theoretically stable definition
of the top-quark mass has been achieved with the introduction of the
threshold mass, which is extracted from a lineshape measurement of
$\sigma_{t\bar t}(e^+ e^- \to t\bar t)$ at threshold (see, e.g.,
Ref.~\cite{Hoang:2006pc} for a review).  $\sigma_{t \bar t}(e^+ e^-
\to t\bar t)$ is known at NNLO QCD~\cite{Hoang:2000yr} and a
theoretically stable extraction of threshold masses with a theoretical
uncertainty of $\Delta m_t =0.1$ GeV is possible~\cite{Hoang:2000yr}.
The relation between the threshold masses and the $\overline{\rm MS}$
mass is known at higher order QCD which enables an extraction of the
$\overline{\rm MS}$ top-quark mass with comparable
precision~\cite{Hoang:2000yr}. While at the ILC a precise top-quark
mass can be obtained from a measurement where only color singlet
$t\bar t$ events need to be counted, the current precise measurement
of $m_t$ at the Tevatron relies on the reconstruction of the invariant
mass of a single top quark. This reconstructed mass is usually
identified with the pole mass which is used in global EW fits. It is
not obvious, if this is justified after non-perturbative effects such
as hadronization, color reconnection, and underlying event modeling
have been taken into account. Moreover, as has been illustrated in the
case of a $m_t$ measurement at the ILC from a $t\bar t$ threshold
scan~\cite{Hoang:2000yr}, when including higher-order corrections, the
pole mass is not stable, since it receives relatively large radiative
corrections from low energy scale physics (``renormalon
problem''). Therefore, it is important to explore alternative methods
and observables that are sensitive to $m_t$ such as $\sigma_{t \bar
t}$ and the invariant-mass distribution of the top-quark pair
($M_{t\bar t}$). A value for $m_t$ has been extracted from the
$\sigma_{t\bar t}$ measurement at the Tevatron with a relative
uncertainty of 4\%~\cite{Abazov:2008gc}. In
Ref.~\cite{Frederix:2007gi}, the theoretical uncertainties of $m_t$
extracted from $\sigma_{t\bar t}$ at NLO QCD are estimated to be
$\Delta m_t/m_t \approx 0.2 \Delta \sigma/\sigma + 0.03$ (LHC),
$\Delta m_t/m_t \approx 0.2 \Delta \sigma/\sigma + 0.016$ (Tevatron),
and $\Delta m_t/m_t \approx 1.2 \Delta M_{t\bar t}/M_{t\bar t} +
0.003$ when extracted from the mean of the $M_{t\bar t}$ distribution
at NLO QCD at the LHC.  At such high precision non-perturbative
effects may become important. For instance, a study of different
underlying event modeling at the Tevatron with Phythia finds $\Delta
m_t =0.5$ GeV~\cite{Skands:2007zg}.  At the LHC, bound-state and
initial-state radiation effects (both calculated at NLO QCD)
significantly modify $d\sigma/dM_{t\bar t}$ at the $t \bar t$
threshold~\cite{Hagiwara:2008df}.  Finally, in
Ref.~\cite{Fleming:2007xt} it has been demonstrated that at least at
the ILC the extraction of a top-jet mass $m_J$ from invariant top-mass
distributions $d^2\sigma/dM_t^2/dM_{\bar t}^2$ ($e^+e^- \to t\bar t$)
in the peak region with better than ${\cal O}(\Lambda_{{\rm QCD}})$
precision seems to be feasible.  In Ref.~\cite{Fleming:2007xt},
$d^2\sigma/dM_t^2/dM_{\bar t}^2$ is calculated in the limit $Q\gg m\gg
\Gamma_t > \Lambda_{{\rm QCD}}$ at NLO QCD with NLL resummation, a
theoretically stable mass definition is used ($m_J$), and a relation
between $m_J$ and $\overline{{\rm MS}}$ and threshold masses is
derived. First steps towards a NNLO and NNLL prediction of 
$d^2\sigma/dM_t^2/dM_{\bar t}^2$ together with a derivation of the relation between $m_J$ and
the $\overline{{\rm MS}}$ mass at two-loop order have been carried out recently in Ref.~\cite{Jain:2008gb}. 
A possible application to a well-defined $m_t$ extraction
from invariant top-mass distributions at hadron colliders is presently
under investigation~\cite{Hoang:2008xm}.

\begin{figure*}[t]
\centering
\includegraphics[height=5.35cm]{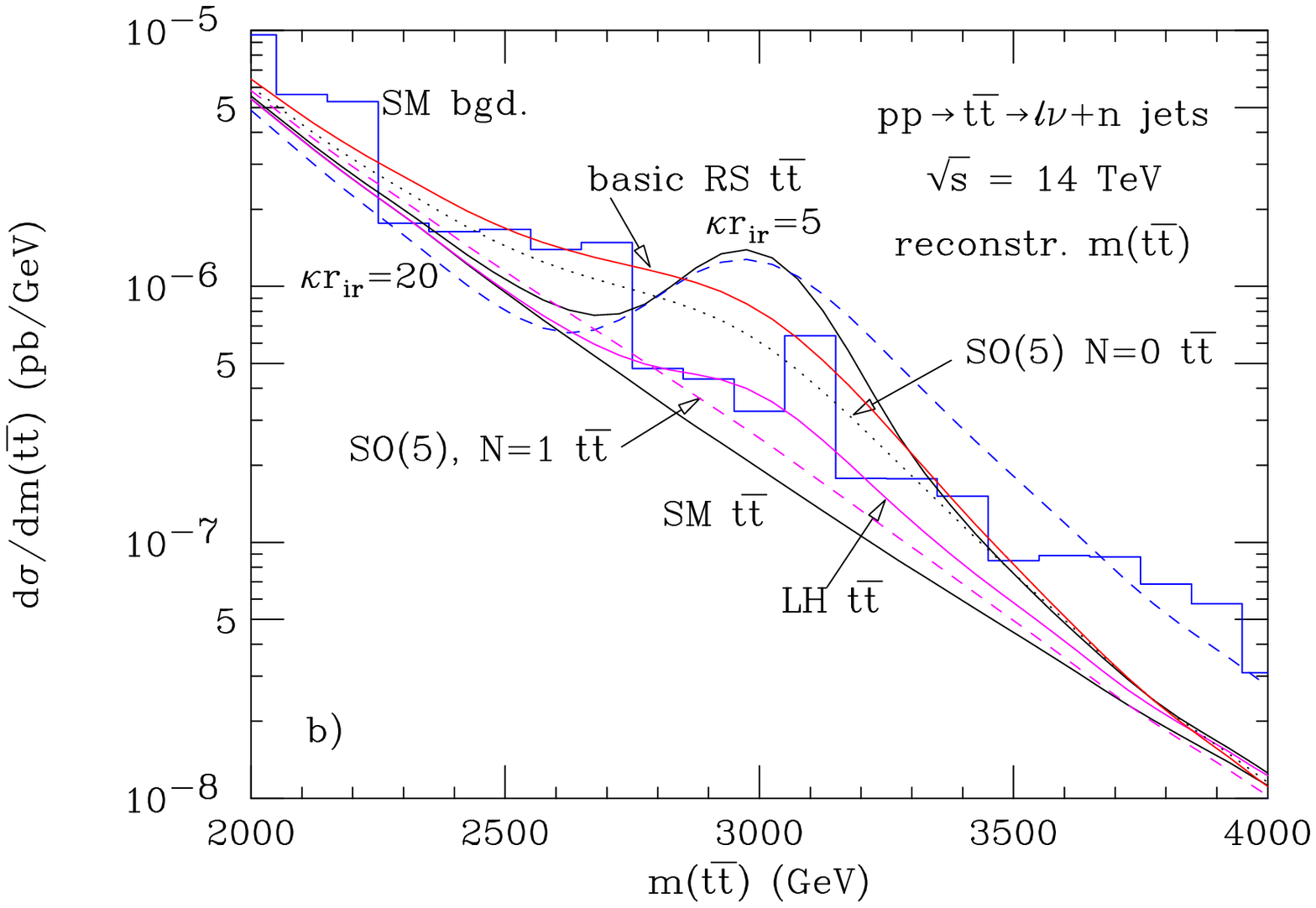}
\includegraphics[width=5.25cm,angle=90]{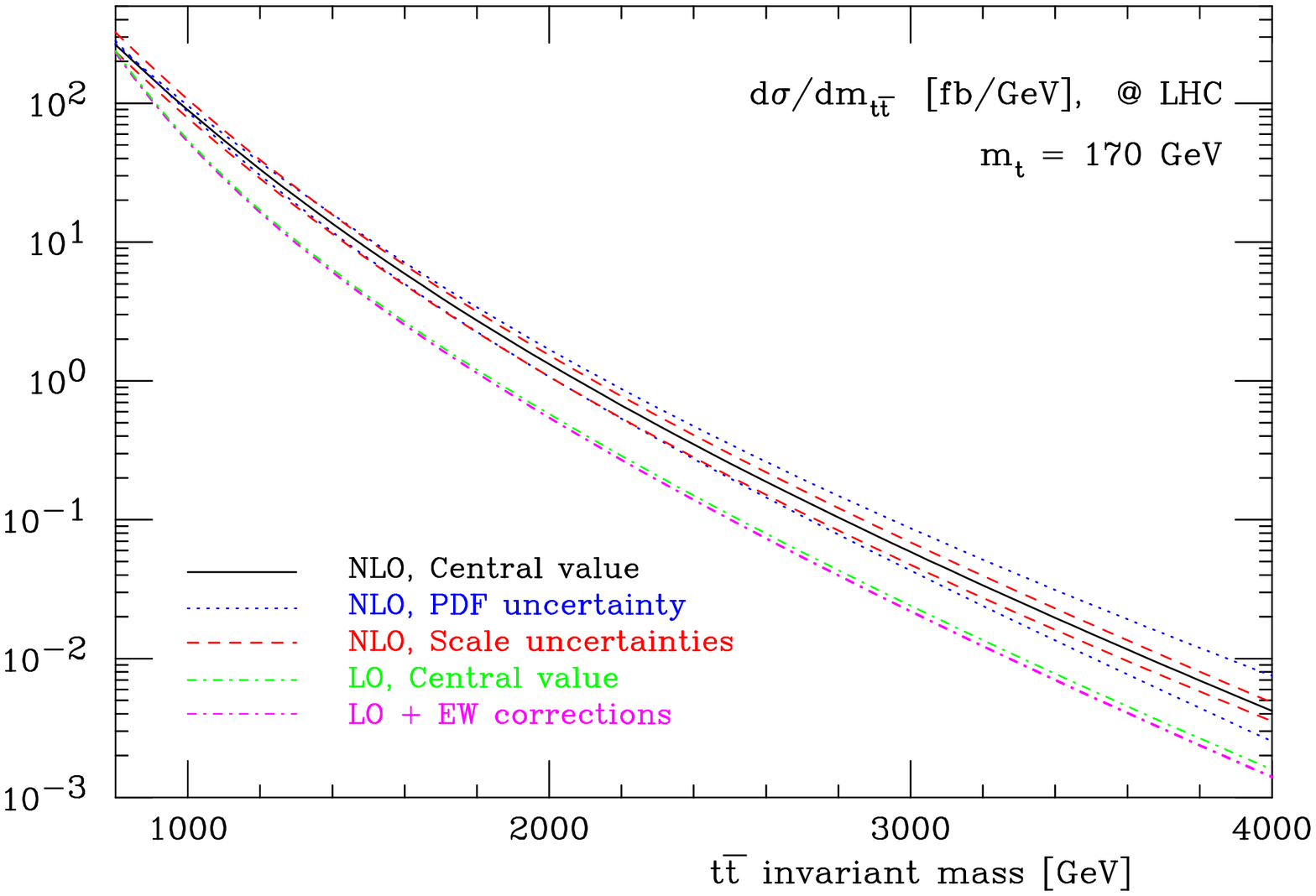}
\caption{The impact of new resonances (l.h.s), i.e. Kaluza-Klein gluons and a $Z_H$ boson in
  Little Higgs models  on the $t\bar t$ invariant mass distribution at high invariant masses at
the LHC, compared with the SM background (taken from
Ref.~\cite{Baur:2008uv}). The r.h.s. plot shows the scale and PDF
  uncertainties of a NLO QCD prediction of $d\sigma/dM_{t\bar t}$ in the same
  kinematic region as well as the LO result (normalized to the NLO total cross
  section) and the impact of NLO EW 
  corrections (taken from Ref.~\cite{Frederix:2007gi}).} \label{fig:mtttail}
\end{figure*}

\section{TOP QUARK DYNAMICS}\label{sec:dynamics} 

The determination of top-quark properties (mass, spin, charge),
searches for signals of non-SM physics, such as anomalous couplings,
FCNC, CP and parity-violating interactions, new $t\bar t$ resonances,
and background studies to Higgs and new physics searches, require good
theoretical control of kinematic distributions in top-quark production
and decay. Figure~\ref{fig:mtttail}, for instance, illustrates the
importance of controlling backgrounds~\cite{Baur:2008uv} and
theoretical uncertainties~\cite{Frederix:2007gi} in the tail of the
$M_{t\bar t}$ distribution in the search of new $t\bar t$ resonances
at the LHC.  For inclusive observables, such as $\sigma_{t\bar t}$,
and $M_{t\bar t}, p_T$ distributions, predictions with on-shell top
quarks usually are sufficient. But for the study of spin correlations
and polarization asymmetries the top-quark decay needs to be
included. The NLO QCD corrections to $t\bar t$ production and decay at
hadron colliders have been calculated by using the narrow-width
approximation including spin correlations between the $t$ and $\bar
t$~\cite{Bernreuther:2001rq,Bernreuther:2004jv}. Asymmetries are
especially interesting tools in the search for non-SM physics, since
they are usually small in the SM.  For instance, within the SM the
forward-backward charge asymmetry in $t\bar t$ production, recently
measured at the Tevatron~\cite{Abazov:2007qb}, is zero at tree-level
and small ($\approx +5 \%$)~\cite{Kuhn:1998jr,Kuhn:1998kw} when
induced by QCD interference effects between initial and final-state
gluon radiation.  The large theoretical uncertainty ($\approx
30\%$)~\cite{Kuhn:1998jr,Kuhn:1998kw} of this prediction is
considerably reduced when NLL threshold resummation of soft gluon
radiation is taken into account~\cite{Almeida:2008ug}.  In $t\bar
t+$jet production at the Tevatron the NLO QCD corrections reduce the
forward-backward charge asymmetry from $-7\%$ to $\-1.5 \pm 1.5
\%$~\cite{Dittmaier:2007wz}.  At the Tevatron, the impact of non-SM
physics, such as tree-level axial couplings of the gluon, can result
in charge asymmetries as large as $-13\%$~\cite{Antunano:2007da}, for
instance.  Also parity-violating asymmetries in the production of left
and right-handed top quarks have the potential to provide a clean
signal of non-SM physics: QCD preserves parity and the SM induced
asymmetries are too small to be observable, at least at the Tevatron
$p \overline{p}$ collider~\cite{Kao:1999kj,Bernreuther:2006vg}.  The
produced top quarks decay almost entirely into a bottom quark and a
$W$ boson before they can hadronize~\cite{Bigi:1986jk} or flip their
spins. The spin correlation of the top-pair system will therefore be
preserved and can be measured by studying angular distributions of the
decay products~\cite{Kuhn:1983ix}.  To measure spin correlations and
asymmetries at the Tevatron and the LHC, higher-order corrections to
polarized $t\bar t$ production need to be known as well: The NLO QCD
and EW corrections to polarized $t\bar t$ production have been
calculated in
Refs.~\cite{Bernreuther:2000yn,Bernreuther:2001rq,Bernreuther:2001bx,Bernreuther:2004jv}
and
Refs.~\cite{Kao:1994rn,Kao:1997bs,Kao:1999kj,Bernreuther:2005is,Bernreuther:2006vg},
respectively. The effects of SUSY QCD and SUSY EW one-loop corrections
to polarized $t \bar t$ production at hadron colliders have been
studied in Ref.~\cite{Li:1997gh,Berge:2007dz} and Ref.~\cite{Li:1997ae,Kao:1999kj},
respectively. They have been found to be promising, but further more
realistic studies are needed, including top-quark decays, in order to
determine whether these effects will be observable at the LHC.

\section{CONCLUSIONS}

There has been a lot of activity and many advances in every aspect of
accurately predicting and modeling top-quark observables at hadron and
$e^+ e^-$ colliders of which only a small subset could be presented
here.  Observables in top-pair and single-top production at hadron
colliders are known at NLO QCD and NLO EW both within the SM and the
minimal supersymmetric SM, and resummation techniques have been
successfully employed to deal with logarithmic enhanced corrections. A
complete fixed-order NNLO QCD calculation for $\sigma_{t \bar t}$ is
needed and the calculation of many of the necessary building blocks
has been completed.  Theoretical uncertainties of the available 
predictions have been studied mostly for inclusive observables
($\sigma_{t\bar t}, \sigma_t \ldots$) and are under investigation for
kinematic distributions.  Theoretically stable top-mass definitions
are needed to match the experimental precision of $m_t$ measurements
at hadron and $e^+e^-$ colliders. The applicability of new results for
a well-defined $m_t$ extraction at the ILC to the LHC is under
investigation and the potential of alternative observables for a
precise $m_t$ measurements at hadron colliders, such as $\sigma_{t\bar
t}$ and $d\sigma/dM_{t\bar t}$, are being studied. Theoretically
promising searches of non-SM signals in asymmetries in $t\bar t$
production, such as the forward-backward charge asymmetry, parity
violating asymmetries in polarized $t\bar t$ production, and spin
correlations between the $t$ and $\bar t$, require the inclusion of
radiative corrections to top production and decay and more detailed
and realistic studies of how to disentangle non-SM signals from SM
backgrounds, for instance.  These are just a few examples of the
theoretical challenges that need to be met in order to fully exploit
the potential of the Tevatron and the LHC for precision top-quark
studies within the SM and beyond.

\begin{acknowledgments}
Work of D.~W.~supported by National Science Foundation grant no.~NSF-PHY-0456681 and no.~NSF-PHY-0547564.
\end{acknowledgments}

\end{document}